\documentclass[10pt,conference]{IEEEtran}
\label{doc:imports}

\usepackage{algorithmic}
\usepackage{microtype}
\usepackage{graphicx}
\usepackage{textcomp}
\usepackage{bmpsize}
\usepackage{xcolor}
\usepackage{lipsum}
\usepackage{pifont}
\usepackage{booktabs}   %% For formal tables: http://ctan.org/pkg/booktabs
\usepackage{subcaption} %% For complex figures with subfigures/subcaptions: http://ctan.org/pkg/subcaption
\usepackage{multirow}
\usepackage{listings}
\usepackage{array}
\usepackage[colorlinks=false, hidelinks]{hyperref}
\usepackage[numbers,sort&compress,square]{natbib}
\usepackage{paralist}
\usepackage{multicol}
\usepackage{rotating}
\usepackage{mdframed}
\usepackage{url}
\usepackage[hang, flushmargin]{footmisc} 
\usepackage[T1]{fontenc}
\usepackage[inline]{enumitem}

\label{doc:settings}
\newsavebox{\fmbox}
\newenvironment{fmpage}[1]{\begin{lrbox}{\fmbox}\begin{minipage}[c][1.4cm]{#1}}
		{\end{minipage}\end{lrbox}\fbox{\usebox{\fmbox}}}

\definecolor{aqua}{rgb}{0.2,0.5,0.9}
\definecolor{paleaqua}{rgb}{0.74, 0.83, 0.9}
\definecolor{whitesmoke}{rgb}{0.9, 0.93, 0.96}
\definecolor{deepblue}{rgb}{0,0,0.5}
\definecolor{deepred}{rgb}{0.9,0.1,0.1}
\definecolor{deepgreen}{rgb}{0,0.5,0}
\definecolor{bittersweet}{rgb}{1.0, 0.44, 0.37}
\definecolor{caribbeangreen}{rgb}{0.0, 0.8, 0.6}
\definecolor{cottoncandy}{rgb}{1.0, 0.74, 0.85}

\lstdefinelanguage{JavaScript}{
	keywords={function, return, if, else, for, while, var, this, new},
	morecomment=[l]{//},
	morecomment=[s]{/*}{*/},
	morestring=[b]',
	morestring=[b]",
	moredelim=[is][{\itshape\color{blue}}]{\#}{\#},
	ndkeywords={},
	keywordstyle=\color{deepblue},
	ndkeywordstyle=\color{blue},
	identifierstyle=\color{black},
	commentstyle=\color{deepgreen},
	stringstyle=\color{deepred},
	sensitive=true
}

\label{doc:settings:newcommands}
\newcommand{\ttcode}[1]{\textcolor{darkgray}{\ttfamily\small{#1}\normalfont\normalsize}}
\newcommand{\myparagraph}[1]{\paragraph*{\textbf{#1}}}

\newcommand{\CONTENTBOX}[1]{\noindent\textcolor{black}{\fcolorbox{whitesmoke}{whitesmoke}{\begin{minipage}[t]{.97\columnwidth}\textit{#1}\end{minipage}\vspace{1em}}}}

\newcommand{\NRTotalPackages}{703,457}
\newcommand{\NRValidPackages}{604,159}

\newcommand{\AttackSurfReduction}{{31.9\%}}
\newcommand{\NoPermissionsPackages}{{192,585}}

\newcommand{\NrApplicationsSampled}{20}

\newcommand{\DatePackagesSampled}{February 2018}
\newcommand{\NrPermissions}{four}
\begin{document}

%%
%% The "title" command has an optional parameter,
%% allowing the author to define a "short title" to be used in page headers.
\title{Containing Malicious Package Updates in \emph{npm} with a Lightweight Permission System}
\author{\IEEEauthorblockN{Gabriel Ferreira, Limin Jia, Joshua Sunshine, Christian K{\"a}stner}
\IEEEauthorblockA{\textit{Carnegie Mellon University}}
}
\maketitle

\begin{abstract}
	The large amount of third-party packages available in fast-moving software ecosystems, such as Node.js/npm, enables attackers to compromise applications by pushing malicious updates to their package dependencies. Studying the npm repository, we observed that many packages in the npm repository that are used in Node.js applications perform only \emph{simple} computations and do not need access to filesystem or network APIs. This offers the opportunity to enforce least-privilege design per package, protecting applications and package dependencies from malicious updates. We propose a lightweight permission system that protects Node.js applications by enforcing package permissions at runtime. We discuss the design space of solutions and show that our system makes a large number of packages much harder to be exploited, almost for free.
\end{abstract}


\section{Introduction}
\label{sec:intro} 
Modern software applications are commonly built on top of many reusable packages that are constantly evolving~\cite{bogart:2016BreakingAPIs,decan2019empirical,manikas2016revisiting}, which raises a risk of supply-chain attacks through \emph{malicious packages updates}. Such kind of attacks target applications or its users, but are performed through updates in applications' package dependencies, which are downloaded into an application automatically or manually by unsuspecting developers. 
\looseness=-1

The risk from malicious package updates, beyond transport security \cite{cappos:2008:attacksOnPkgManagers, samuel:2010:survivableKeyCompromise, nikitin:2017:proactiveSoftUpdate}, has long been ignored or seen as a theoretical possibility only \cite{npm-hydra-worm-disclosure, gilbertson:2018:harvesting}. However, recently, more and more cases of malicious package updates have been discovered in multiple large open-source repositories \cite{eslint:2018:postmortem, getcookies:2018:postmortem, event-stream:2018:postmortem, electron-native-notify:2019:postmortem, maliciousPackagesRuby, maliciousPackagesPython}. Attackers keep finding ways to obtain control of developer accounts (e.g., using leaked credentials, targeting weak passwords, or offering help to maintain a package). When in control of an account, attackers can publish a modified malicious version of the package, which is then downloaded (often automatically) by applications depending on this package. Figure~\ref{fig:eslint-attack} shows an excerpt of a real attack.

\begin{figure}
	\begin{subfigure}{\linewidth}
		% // imports "https", downloads and executes script 
		\begin{lstlisting}[caption={}]
var https = require("https");
https.get({	hostname: "pastebin.com", path: "/evil" }, 
	r => { r.on("data", c => { eval(c); }); }
).on("error", () => {});
		\end{lstlisting}
		\vskip -1em
		\caption{After a malicious update, the package now downloads and executes the script below.}
	\end{subfigure}	
	\begin{subfigure}{\linewidth}
		\noindent
		\vskip -.5em
		% // imports "fs", reads and leaks npm credentials file
		\begin{lstlisting}[caption={}, frame=tlrb, label={code:eslint-pastebin}]
var fs = require("fs");
var npmrc = require("path").join(...,".npmrc");
if (fs.existsSync(npmrc)) {
	var content = fs.readFileSync(npmrc, "utf8")
	var https = require("https");
	https.get({ hostname: "evil.com", method: "GET", 
		headers: {Referer: "http://1.a/"+content}},()=>{}
	).on("error",()=>{}); 
}
		\end{lstlisting}
		\vskip -1em
		\caption{Downloaded malicious script reads and leaks \emph{npm} package manager credentials.}
	\end{subfigure}	
	\caption{Essence of the \textbf{eslint-scope@3.7.2} attack.}
	\label{fig:eslint-attack}
	\vspace{-.4em}
\end{figure}

While malicious package updates are a potential problem in all software projects with external dependencies, we will
argue that common practices and design decisions in the \emph{Node.js/npm}
ecosystem make such JavaScript applications a particularly attractive target for
malicious package updates. Among others (see Sec.~\ref{sec:characteristics} for details),
they tend to depend on many small external libraries, 
they tend to allow automatic updates of minor updates,
and the runtime gives all packages the same application-level privileges.
When faced with numerous updates from many direct and indirect dependencies,
\emph{Node.js/npm} developers often enable automated updates, despite potential security risks.

Many defenses against supply-chain attacks have been developed ~\cite{npmReviews, npm2fa, npmAudit, snyk, jslint, meyerovich:2010:ConScript, shen2018rescue, garret:2019:detectingSuspiciousPackageUpdates, cappos:2008:attacksOnPkgManagers, samuel:2010:survivableKeyCompromise, nikitin:2017:proactiveSoftUpdate}, but they tend not to be practical in many \emph{realistic software engineering settings}.
Defenses include carefully reviewing all dependencies and dependency updates~\cite{npmReviews},
hardening the package infrastructure (e.g., transport security, two-factor authentication)~\cite{cappos:2008:attacksOnPkgManagers, samuel:2010:survivableKeyCompromise, nikitin:2017:proactiveSoftUpdate, npm2fa}, and various forms of program analysis and anomaly detection~\cite{jslint,meyerovich:2010:ConScript,shen2018rescue,garret:2019:detectingSuspiciousPackageUpdates,snyk, npmAudit}.
However, as we will discuss, in a practical software engineering perspective current approaches either (a) are too expensive for practical
use, (b) require a complete redesign of the Node.js module system or runtime environment that is unlikely to see adoption in practice,
or (c) only defend against already known vulnerabilities.
\looseness=-1

In this paper, we design a lightweight permission system and a corresponding enforcement mechanism that protects applications against malicious updates from a large number of packages in direct and indirect package dependencies. 
Our solution is partial, in that it only defends against attacks of
a subset of packages, but it is explicitly {designed} to be \emph{easy to adopt} and has \emph{negligible runtime
overhead}, making it an important and practical building block in defending against
malicious package updates.

We build on the insight that \emph{many Node.js packages perform simple computations and do not need access to security-relevant resources, such as the filesystem or the network APIs or metaprogramming constructs}.
Our solution effectively sandboxes the large number of simple third-party packages
in the \emph{Node.js/npm} ecosystem that do not require access to
security-critical resources, making malicious updates attacks that attempt to elevate packages' privileges ineffective.

The novelty of our permission system lies in the \emph{design} of a practical and lightweight solution that
focuses on providing useful and easy to adopt, albeit partial protections.
Where existing sandboxing solutions require invasive changes to infrastructure
or package implementations, or impose severe
runtime overhead \cite{deGroef:2014nodeSentry, tran:2015:JaTE, breakapp:ndss:2018}, ours integrates with the current \emph{Node.js} infrastructure
without changes to the implementation of existing packages and imposes negligible runtime overhead. 
\looseness=-1
Even though we cannot protect all packages,
% In the \emph{design space} of possible security techniques, we propose
% a technique that can provide strong protections for a subset of 
% packages almost for free.
% While packages that legitimately need access to security-relevant resources to deliver their functionality cannot be protected with our approach,
% protecting simpler packages means that developers and security expert
% can focus their attention more deliberately on packages our technique does not protect.
taking a software engineer's system perspective in the fast paced world of open-source software ecosystems, we argue that even a 10~percent reduction in attack surface that can actually be enacted broadly would result in significant saving of community resources for security reviews and would make it harder for attackers to find packages that they can exploit. 
\looseness=-1

% We implement and evaluate our design for a permission system for \emph{Node.js/npm}. 
Our evaluation shows that \AttackSurfReduction\ of all \emph{npm} packages can be protected by our design and that 52~percent of one year's package updates in 120~popular npm packages and applications are for those protected packages. In addition, our implementation's average performance overhead is negligible  ($\ll 1\%$).
\looseness=-1

Overall, we make the following contributions: (1) we design a lightweight permission system that protects Node.js applications against malicious package updates for a significant number of packages, (2) we discuss design trade-offs to highlight how the chosen partial but low-cost solution fits into a larger security strategy, (3) we evaluate the solution on a large number of packages and applications, and (4) we make both the implementation and evaluation benchmarks available (\href{https://github.com/gabrielcsf/malicious-updates-icse2021}{\footnotesize\ttfamily https://github.com/gabrielcsf/malicious-updates-icse2021}).

\section{The Problem: Malicious Updates, npm, and Current Defenses}
\label{sec:overview}

We focus on malicious package updates in the Node.js/npm ecosystem, which is the largest, most popular, and fastest growing open-source ecosystem with over one million reusable packages available to download. Several actual attacks were found recently (discussed below), emphasizing the importance of the problem.

\myparagraph{Node.js/npm}
To explain the problem and our solution, it is important to understand
how packages and updates work in Node.js/npm. Node.js is a runtime system that provides powerful APIs to interact with 
the host system (files, network, processes), which enable programmers to write applications beyond JavaScript's traditional use in a browser. 
While early applications were heavily biased toward backend web servers,
Node.js is also popular for command-line, desktop, robotics, and IoT applications.
 
Node.js provides its own module system, where each JavaScript file is loaded as a module. Once loaded, modules are represented as JavaScript objects. Node.js projects are structured into modules, which are grouped into named packages or applications. Core APIs are offered through a small set of \emph{native modules}, but developers routinely import a large number of additional modules from third-party packages. Besides JavaScript files, a package contains a manifest file that lists package dependencies required for it to work properly.

\looseness=-1
Node.js is tightly integrated with \emph{npm}, a package manager and a repository for Node.js packages. The package manager \emph{npm} provides convenient mechanisms to download, install, and update packages and their recursive dependencies from the \emph{npm} repository. In a typical applications's (or package's) installation process, the package manager interprets the content of the manifest file, resolves packages versions, and downloads the source code of direct and indirect packages listed as dependencies.

The design of the \emph{npm} package manager encourages automatic updates and favors ease of publishing packages~\cite{bogart:2016BreakingAPIs}. Package dependencies can be pinned down to specific versions or defined as version ranges~\cite{preston-werner:2018:semver,decan2018evolution,dietrich2019dependency}; the use of ranges to automatically install minor updates is very common~\cite{hejderup:2015:masterThesis,zerouali2018empirical,cogo2019empirical,dietrich2019dependency}. 

\myparagraph{Node.js/npm's characteristics facilitate malicious update attacks}
\label{sec:characteristics}
Attacks through malicious package updates are possible in most software ecosystems, though certain characteristics make Node.js/npm a particularly attractive target:

\begin{itemize}[leftmargin=*]
	\item The JavaScript language and Node.js platform provide only a small set of native modules and essentially \textbf{no standard library}, leaving it up to the community to develop packages even for standard tasks such as string manipulation and collections. Hence, developers
	often depend on many third-party packages, even for simple functionality, contributing to a \emph{large attack surface}.\footnote{Informally, we consider the number of accounts that can update any of an application's dependencies as the attack surface; the more accounts involved, the higher the chance that any one of them may be compromised.}
	\item The Node.js/npm community prefers a model of \textbf{many small packages} (inspired by the Unix philosophy) \cite{abdalkareem:2017}. Thus, it is common to depend on a large number of packages, where each of those packages contributes to a \emph{large attack surface}.
	\item Developers commonly provide version ranges on dependencies, such that patch-level \textbf{updates are automatically installed}, depending on version labels set by the package maintainer~\cite{hejderup:2015:masterThesis}. 
	% Beyond development and test systems, even some production systems
	% automatically install and update third-party dependencies on a release. 
	The practice 
	of installing updates automatically in development, test, and sometimes even production systems, contributes to making applications \emph{easy to exploit}.
	\item The Node.js/npm community values ease of publishing, where updates can be published with a single command-line instruction (typically with locally stored credentials) without further quality checks or reviews~\cite{bogart:2016BreakingAPIs}. Due to a constant stream of updates, \textbf{developers update frequently}, to avoid having to update many packages across many versions at once~\cite{bogart:2016BreakingAPIs}. This also makes applications \emph{easy to exploit}.
	\item Most packages also have dependencies of their own, so adding a single
	package dependency often comes with \textbf{many indirect package dependencies} that are \textbf{de-facto invisible} to developers. Hence, indirect dependencies are an attractive target for attackers, making applications \emph{easy to exploit}.
	\item Node.js applications are typically deployed as single-threaded applications, in which \textbf{all loaded packages inherit the applications' privileges to use security-relevant resources} from accessing local files and the network, to modifying global objects and other packages, to generating code at runtime \cite{richards:2011}. 
	As a consequence, loaded malicious packages have a \emph{high potential for damage}. 
\end{itemize}

We exemplify the ease of exploit and the potential damage with three recent attacks detected in the last three years:

\begin{itemize}[leftmargin=*]
	\item In 2019, the \emph{npm, Inc.} security team identified and reported a malicious version of the \emph{electron-native-notify} package \cite{electron-native-notify:2019:postmortem}. The attacker published the package with useful functionality and waited until it was added as a dependency of the \textit{Agama} wallet application before
	publishing a malicious update. The attacker stole about \$13 million dollars in bitcoin tokens. 
	\item Also in 2019, the popular \emph{event-stream} package was updated maliciously to steal bitcoins~\cite{event-stream:2018:postmortem}. 
	The malicious update was discovered only after 2.5 month and 8 million downloads. The original maintainer of the \emph{event-stream} package had handed over the account to the attacker when he offered to help maintain the project
\looseness=-1
	 (i.e., social engineering). 
	\item In 2018, the \emph{eslint-scope} package, part of a widely used JavaScript linter, was also a target of a malicious update. The attack aimed at stealing the \emph{npm} package manager credentials from users of the linter and affected around 4500 accounts (see Figure~\ref{fig:eslint-attack}).

\end{itemize}

\myparagraph{State-of-art defenses}\label{sec:stateoftheart-defense}
In current practice in Node.js/npm, a number of strategies can
lower the risk from malicious package updates, though all have severe limitations:
\begin{itemize}[leftmargin=*]
	\item \emph{Inspection:} Node.js developers are unlikely to carefully audit the large number of direct and indirect dependencies and their updates. Developers typically hope that the community at large will
	find and report vulnerabilities quickly% (``many eyes will make all bugs shallow'' hypothesis)
	, but past attacks remained undetected for months or caused significant damage within short periods.
	Current static analysis and anomaly detection tools detect usually only very specific issues and produce many false alarms~\cite{jslint, zampetti2017openSA, shen2018rescue, garret:2019:detectingSuspiciousPackageUpdates}.

	\item \emph{Tracking known vulnerabilities:} Many third-party services
	scan the dependency tree of Node.js applications for known vulnerabilities (e.g., \emph{Snyk.io}, \emph{npm}, \emph{GitHub}). This strategy is reactive, and research has shown that developers are developing
	notification fatigue and are slow to update~\cite{bogart:2016BreakingAPIs, decan:2017, wittern:2016, derr:2017:KMU, mirhosseini:2017:APR, kula:2018:doDevUpdateDep}.

	 \item \emph{Avoiding automatic updates:} Rather than using automatic
	 updates with version ranges, developers may \emph{lock} package versions
	 or use bots to only update dependencies after executing tests~\cite{greenkeeer,renovate,mirhosseini:2017:APR}. However, it is not clear that  automated test executions would detect malicious updates.
	
	\item \emph{Infrastructure hardening:} two-factor authentication in the \emph{npm} package manager \cite{npm2fa} reduces some attack vectors but does not protect against attacks using social engineering as in past incidents.
	
	\item \emph{Application-level sandboxing:} Some Node.js applications
	are deployed within a sandbox (e.g., containers \cite{nodejsdockernodesource}), reducing potential damage.
	However, sandboxing is done at application level where all packages
	have the same capabilities as the application (where the application often
	rightfully has access to files, databases, or the network).
\end{itemize}

All these practices help but offer only limited protection.
More secure solutions from academic security research on isolating individual packages or tracking
information flows (cf. Sec.~\ref{sec:alternativeDesign}\,\&\,\ref{sec:related-work}) are not adopted in practice because of their
limitations. We complement existing practices with an easy to adopt and low-overhead sandboxing strategy at the package level that can substantially reduce the attack surface.

\section{Permission System Design}
\label{sec:perm-system}

\noindent
We propose a permission system that sandboxes packages and enforces \emph{per-package} permissions in Node.js applications, i.e.,  we enforce a \emph{least-privilege design~\cite{viega2011building} at the package level}.
\looseness=-1

Our approach is not the first to sandbox individual \emph{npm} packages~\cite{deGroef:2014nodeSentry,breakapp:ndss:2018,tran:2015:JaTE} (cf. Sec.~\ref{sec:related-work}), and there is a large design space for possible solutions, as we will discuss in Sec.~\ref{sec:alternativeDesign}. 
However, our approach identifies \emph{a novel
design} that provides protections for a large subset of packages 
without requiring changes to
package implementations and with negligible overhead.
%As a consequence, our design is more lightweight than prior work on isolation and compartmentalization of applications and packages ($\ll 1\%$ rather than $>20\%$ performance overhead in prior approaches \cite{breakapp:ndss:2018, deGroef:2014nodeSentry, tran:2015:JaTE}).
We align our design with the requirements and values of  the Node.js/npm community and propose it as one useful building block in a security strategy.
\looseness=-1

\subsection{Goals and Assumptions}
\label{sec:goalsAndAssumptions}
The design of the permission system focused on three main goals that are important for it to be relevant in practical software engineering settings: First, the permission system should actually \emph{reduce the attack surface} of applications by containing certain types of attacks. Second, the permission system should \emph{not require major infrastructure changes}, be backward compatible, and \emph{not break existing user code} (assuming sufficient permissions). Lastly, the proposed permission enforcement technique should have \emph{low performance overhead}, which is relevant for practical adoption. 
\looseness=-1

In this work, we focus exclusively on malicious package updates, which are attacks following the following pattern:
First, an attacker obtains credentials of package developers by using leaked \emph{npm} package manager credentials in Git repositories,
gaining access to a package developer's machine, 
buying packages, or using traditional tactics such as targeting weak passwords, phishing, social engineering, and typo-squatting.
Note, it is sufficient to compromise the credentials of a single developer
among an application's often hundreds of transitive dependencies. 
Second, once the attacker uses the credentials to publish malicious code with an update, applications that directly or indirectly depend upon the package and install updates (automatically or manually) are at risk.
Once a malicious package is loaded in a running application, it may import native modules, import modules from other packages, and use metaprogramming constructs to perform malicious actions (see Fig. \ref{fig:eslint-attack}).\looseness=-1
%. For example, an attacker can combine the \emph{http} module and the \emph{eval} function to leak data over the network 

\subsection{Package Permissions}
\label{sec:permissions}
We follow a familiar permission strategy,
as known from mobile apps or web-browsers extensions:
(1) developers declare required permissions from a small
set of common and easy to understand permissions for their packages,
which would be shown in the \emph{npm} repository and by the command-line
tools,
(2) the system enforces that the package does not use not-required permissions,
and (3) developers who add or update a package dependency must accept the package's
permissions at installation time and again when permissions change in an update.
\looseness=-1

\myparagraph{On permission systems}
These kinds of permissions systems are well understood by users, easy
to use for developers, and also well studied, including problems of developers asking for too many permissions,
and users ignoring permissions~\cite{bartel:2012:reducingAttackSurfaceAndroid, wei:2012:permEvolutionAndroid, felt:2011:androidPermDemystified, guha:2011:verifBrowserExt}.
Our design shares similar challenges, but we expect
fewer practical problems due to fewer monetization concerns
(e.g., many Android permissions are needed just for targeted advertising)
and a different target audience:
Package users are developers and can usually clearly understand why
a package would or would not need specific permissions
(e.g., a string template engine needing network access would raise
immediate suspicion).
As permission changes on Node.js/npm are rare and suspicious, especially for minor and patch updates (see Sec.~\ref{sec:evaluation-rq4}), developers and the community at large are
much more likely to focus their attention on such updates.
\looseness=-1

\myparagraph{Set of permissions}
Our design is not limited to a specific set of permissions (i.e., other specific permissions can be defined and mapped to other security-relevant resources, if desired), but for our discussion, implementation, and evaluation we consider \NrPermissions{} easy to understand permissions:
\begin{itemize}[leftmargin=*]
\item The \textbf{network} permission is required to reference APIs to communicate with remote servers (e.g., HTTP, sockets). Specifically, the  native modules \emph{http, http2, https}, and \emph{net} require this permission. Without the \emph{network} permission, malicious code cannot leak data over the network.

\item The \textbf{filesystem} permission is required to reference APIs to access the local filesystem, especially the  native module \emph{fs}. Without this permission code cannot perform attacks that read, write, or delete local files.

\item The \textbf{process} permission is required to reference APIs for interacting with operating-system processes, particularly the native module \emph{child\_process}. Without this permission, malicious code cannot open reverse shells or kill processes.

\item The \textbf{all} permission is required to use metaprogramming constructs (e.g., \emph{eval, with}). Without this permission, malicious code cannot affect applications globally (e.g., modify the prototypes of native objects) and cannot evade the permission system. The \textbf{all} permission is a \emph{superset} of the other permissions, since the use of metaprogramming constructs enables packages to obtain references to the security-relevant resources enabled by the other three permissions.

\end{itemize}
\looseness=-1

In our specific design, each package may require one or multiple permissions, which then apply to all modules in that package.
Intuitively, the code of a package can only import modules from packages that have the same or fewer permissions. Permissions of native modules are hard-coded.

As we will explain, mapping permissions to code,
a source object from a module with permissions $X$ 
may only hold a reference to a value originating from a module with
permissions $Y$ if $Y\subseteq X$.
\looseness=-1

\myparagraph{Package permissions are composed transitively}
To depend on another package, a package must have at least
the same permissions. For packages, this implies that they cannot 
circumvent the permission system by delegating
critical tasks to packages that have the suitable permissions.

For developers, this means that they can easily (i) interpret a package's permissions without also investigating all indirect dependencies and (ii) declare the needed permissions for their own packages based on which permissions imported packages need.

\myparagraph{Permission enforcement}
\label{sec:perm-system-design}
The challenge when designing a permission system is in enforcing permissions, to prevent that attackers gain access
to resources for which a package does not have permission. Intuitively, our enforcement mechanism needs to ensure that \emph{source} objects from a module cannot import \emph{target} objects from another module, including native modules, for which they do not have permission to import.

We considered different enforcement design options, but settled
on a lightweight sandboxing strategy that combines dynamic checks with static analysis.  We arrived at this design after exploring alternative designs in the design space for a practical and lightweight, yet effective solution; we discuss alternative designs and their trade-offs in Sec.~\ref{sec:alternativeDesign}.

\subsection{Specification: Protecting Security-Relevant Resources}
\label{sec:specification-policy}
Given the dynamic nature of JavaScript and the design of Node.js, there are many ways code can gain references to objects from other modules.
In Figure~\ref{fig:policy}, we define a concrete
policy that our permission system aims to enforce:
If a module has the \textbf{all} permission, we do not enforce any restrictions (case 1),
otherwise we only allow references to objects from modules with
the same or fewer permissions, typically received via import (case 2), with the \emph{global} object being
a special case that may always be referenced (case 3).
In addition, we allow three recursive mechanisms from which modules can derive new
references from legally held references: 
received as arguments from a function call where the caller was allowed
to hold the reference (case 4), received as return value
from a call to a legally referenced function (case 5), or received by accessing the properties of a legally referenced object with
the exception of a few restricted properties (case 6).

More intuitively, the specification prohibits \emph{actively} importing 
objects from modules without suitable permissions, but it
allows code to receive and hold references without corresponding permissions
when those references are explicitly \emph{provided by other modules} through
function parameters, return values of function calls, or global variables.
This design allows modules to pass objects (including callback
functions) to modules with fewer permissions.
It is the caller's responsibility not to provide
security-critical references to untrusted code and it is unlikely
that a malicious package update for a package without permissions can 
expect being passed the right security-critical resource (e.g., the \emph{http} module) as an argument. This restriction puts some burden on developers, but is standard in the design of permission and effect systems~\cite{melicher2017capability}.

In case~6, we restrict the following properties from the \emph{require}, \emph{module}, and \emph{global} objects: 
\ttcode{\{require.main, module.paths, module.\_load, module.globalPaths, module.constructor, module.parent, module.children, global.eval, global.Function, global.process\}}. In addition, we restrict \ttcode{\{prototype, \_\_proto\_\_, create, setPrototypeOf\}} for \emph{native objects} (e.g., \emph{Object}). 
All of these may lead to
unprotected import mechanisms or enable non-local changes via
metaprogramming such as prototype pollution attacks~\cite{phung:2009:lightweightselfprotectingJS}; modules rarely use
these access paths in practice, and they can still continue to do
so if needed, requesting the \textbf{all} permission.

\begin{figure}[t]
\small
	\CONTENTBOX{A \textbf{source} object (representing module A) may hold a reference to a value \textbf{v} originating from a module B if and only if:
		\begin{enumerate}[label=(\arabic*)]
			\item package A has the \textbf{all} permission, or
			\item A has at least the same permissions as B, or
			\item \textbf{v} is the \textbf{global} object, or
			\item \textbf{v} was received from a third object as a parameter in a function call to the \textbf{source} object, where the third object may hold a reference to \textbf{v}, or
			\item \textbf{v} is the resulting value of calling a function \textbf{f} and the \textbf{source} object may hold a reference to \textbf{f}, or
			\item \textbf{v} is the value held by property \textbf{p} of an object \textbf{o} where the \textbf{source} object may hold a reference to \textbf{o}, unless property \textbf{p} is restricted for object \textbf{o}
		\end{enumerate}
	}
	% \vspace{-1.2em}
	\caption{Protecting Access To Security-Relevant Resources.}
	\label{fig:policy}
\end{figure}

\subsection{Enforcement: Protecting Security-Relevant Resources}
\label{sec:perm-system-design-enforc}

To prevent {active} access to objects from other modules without suitable permissions we need to effectively perform checks only for two actions:
First, we need to control module imports with Node.js' \emph{require} function and, 
second, we need to control property access for certain
restricted properties (which could be used for accessing other import mechanisms and evading the sandbox).
By working with the existing implementation structures
of Node.js and making small modifications to the runtime,
both restrictions can be enforced without requiring developers to 
modify their packages and with negligible runtime overhead, as we will explain.

% As described in the specification, there are two ways to which a package can \emph{actively} obtain a reference to a security-relevant resource: 
% \begin{inparaenum}[(1)]
% 	\item by importing a package that has the permission to a security-relevant resource (e.g., \textbf{network}), or
% 	\item by referencing special objects or object properties that provide alternative mechanisms or allow packages to evade the permission system.
% \end{inparaenum} Therefore, to enforce our policy, we need
% \begin{inparaenum}[(1)]
% 	\item to mediate the use of \emph{require}, the main import mechanism, by intercepting calls to the \emph{require} function and checking permissions, and
% 	\item to mediate property accesses in special objects, by intercepting attempts to access blacklisted properties and checking permissions.
% \end{inparaenum} 	

\label{sec:intercepting-require}
\myparagraph{(1) Mediating the main import mechanism (\emph{require})} 
Performing additional runtime checks during imports is straightforward
with only small modifications of the Node.js runtime (without any modifications of the modules).

Node.js provides a function \emph{require} to every module that can
be called to import packages (technically using the \emph{Module Pattern} \cite{osmani:2012:javaScriptDesignPatterns}).
This design is beneficial for us, since each module already receives
its own \emph{require} function.
To intercept all imports, we simply wrap the provided \emph{require}
with one that conducts permission checks for this module with a
one-line modification of the Node.js runtime shown in Figure~\ref{fig:enforcement-wrapping} -- thus comparing the permissions of the importing package with those of the imported package.
Beyond the one-line modification to insert the wrapper, our implementation
for loading and comparing permissions is less than 100 lines of code.
\looseness=-1

Note that a module can potentially gain access to a \emph{require} function
of a different module.
To prevent \emph{active} access to other module's require functions,
the access path \emph{module.parent} is restricted.

\begin{figure}
	\begin{lstlisting}[mathescape=true]
Module._compile = function(code, file) {
	#var rcode = propAccessRewrite(code);# $\label{line:rewrite}$
	var wcode = Module.wrap(rcode);
	var cwpr = vm.runInThisContext(wcode);
	var dir = path.dirname(file);
	#var wreq = wrapRequire(require, dir);# $\label{line:require}$
	var args = [this.exports, #wreq#, this, dir];
	return cwpr.apply(this.exports, args);
}
function wrapRequire(require, currentModule) {
	var permA = lazy(loadPermissions(currentModule));
	return function(targetModule) {
		var permB = loadPermissions(targetModule);
		if (!subset(permA(), permB)) 
			throw new Error('...');
		return require(targetModule);
}}
	\end{lstlisting} 
	\vspace{-1em}
	\caption{Package loading mechanism in the Node.js runtime system. Our modifications are highlighted in blue: the re-writing of property accesses (Line~\ref{line:rewrite}) and the wrapping of the \emph{require} function with permission checks (Lines~\ref{line:require}).}
	\label{fig:enforcement-wrapping}
\end{figure}

\label{sec:intercepting-property-access}
\myparagraph{(2) Mediating property accesses in special objects} 
The second part of our enforcement, preventing access to certain restricted properties
for certain objects, requires slightly more extensive changes, but
is also fairly straightforward. 
Since we cannot fully statically reason about property access in JavaScript,
we combine static analysis with selected dynamic checks.
At load-time, we automatically
\emph{rewrite} the code of each module without the \textbf{all} permission
to insert dynamic checks for every property access for which we cannot
statically exclude that it may access a restricted property.

% In the current state of Node.js runtime system, once a package is loaded, it can access any global or local objects (e.g., global, Object, \emph{require} and \emph{module}), or properties of these objects available in their namespace. To prevent a package from referencing properties in these objects (see policy in Sec.~\ref{sec:specification-policy}), and consequently, to prevent a package from being able to evade our enforcement mechanism or to produce global effects in a program, we re-write the source code of packages. We only re-write code of packages without the \textbf{all} permission (which can access any security-relevant resources at runtime).

Our rewrite rules, which we apply to all modules without the \textbf{all}
permission at load time work in two steps: normalizing references to global variables and introducing dynamic checks.
% (1)~\emph{Normalize references to global variables:} 
	First, using scope analysis, we rewrite references to global variables to make the property
	access visible, for example, rewriting \emph{console.log} to \emph{global.console.log} (the \emph{with} statement which may prevent
	accurate scope analysis requires the \textbf{all} permission).
	% (2)~\emph{Introduce dynamic checks:}
	Second, for every property access in the form \ttcode{x.y} or \ttcode{x[y]}  (outside the right-hand side of an assignment),
	we introduce a dynamic check \ttcode{\$\$prop(x, y)}, unless 
	\ttcode{y} can be resolved statically (name or string literal)
	to a name that is not in the list of restricted properties.
	Function \ttcode{\$\$prop} (fresh random name generated for every module)
	checks whether the object-property combination is restricted,
	as shown in Figure~\ref{fig:enforcement-rewriting-prop}.
\looseness=-1

\begin{figure}[t!]
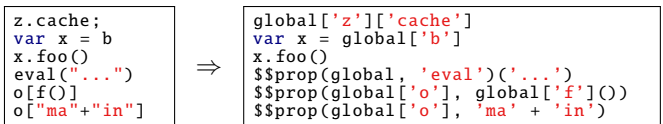

	% \vspace{.5em}
	\begin{fmpage}{.23\columnwidth}
		\begin{lstlisting}[numbers=none,numbersep=0pt,xleftmargin=0cm,]
z.cache;
var x = b
x.foo()
eval("...")
o[f()]
o["ma"+"in"]
\end{lstlisting}
	\end{fmpage}
	\hfill
	$\Rightarrow$
	\hfill
	\begin{fmpage}{.6\columnwidth}
\begin{lstlisting}[numbers=none,numbersep=0pt,xleftmargin=0cm,]
global['z']['cache']
var x = global['b']
x.foo()
$$prop(global, 'eval')('...')
$$prop(global['o'], global['f']())
$$prop(global['o'], 'ma' + 'in')
\end{lstlisting}
	\end{fmpage}
	\vspace{-.5em}
	% \vspace{-1em}
	\caption{Examples of member expressions re-writing (including normalization of global properties).}
	\label{fig:enforcement-rewriting-rules}
\end{figure}

Our rewrite technique \emph{conservatively} combines static and dynamic analysis. For many
property access locations, we can statically identify the property name and avoid
dynamic checks if the name is never restricted. If the name is restricted
for some objects, we insert a runtime check to determine whether the target object
is the restricted one; if we cannot statically resolve the property
name, we introduce a dynamic check, as illustrated in Figure~\ref{fig:enforcement-rewriting-rules}. In practice few dynamic checks are needed
in most modules.
\looseness=-1

\myparagraph{Implementation}
We implement our permission system with the described extensions to 
the Node.js platform. We store declared permissions in a dedicated file in the package.
We use \emph{esprima} to rewrite code and \emph{escope} to analyze scope. 
\looseness=-1
%Both parts are implemented in less than 400 lines of JavaScript code each.

\vspace{.5em}

\begin{figure}[t]
	\begin{lstlisting}
restrictedMap["parent"] = module;
restrictedMap["eval"] = global;
restrictedMap["prototype"] = Object;
...
function $$prop(obj, p) {
	if (obj == undefined) return undefined;
	if (restrictedMap[p] && obj == restrictedMap[p]) 
		throw new Error('...');
	return obj[p];
}
	\end{lstlisting} 
	\vspace{-1em}
	\caption{Simplified code of dynamic checks with \ttcode{\$\$prop}.}
	\label{fig:enforcement-rewriting-prop}
\end{figure}

%\myparagraph{Discussion} 
%Notice how our solution is lightweight, intercepting calls to \emph{require}
%with small changes to the Node.js runtime and dynamically ensuring that certain properties are not referenced. No further expensive information flow
%or origin tracking is needed to enforce the policy at runtime. 
%While more complicated packages cannot be protected, because they genuinely need permissions to provide rich functionality,
%we prefer a lightweight and practical solution for a significant number of packages over a heavyweight and slow solution that supports more flexible policies but that practitioners will not adopt. Given the scale of the \emph{npm} repository and the very large number of dependencies used in many applications, we consider even small security improvements as a positive. Even if our solution can provide protection for only a fraction of all dependencies, we consider this as an important contribution to the security of the Node.js ecosystem, given the low costs imposed by our design.

\section{Design Space}
\label{sec:alternativeDesign}

Notice how our solution is lightweight, intercepting calls to \emph{require} with small changes to the Node.js runtime and dynamically ensuring that certain properties are not referenced. No further expensive information flow or origin tracking is needed to enforce the policy at runtime.

When designing our permission system, both policy and corresponding enforcement system, we explored many design alternatives and settled on the presented
design after multiple experiments and iterations. While our design may appear
simple, limited, or even obvious in retrospect, it is actually the result of
careful consideration of multiple trade-offs, involving provided security
guarantees, flexibility and understandability, 
backward compatibility and ease of adoption, and runtime overhead. 
In the following, we describe the design
space, alternative designs, and justify our design decisions.
Let us start by discussing general concerns before exploring two obvious alternative designs.

\myparagraph{Permission granularity}
We assign permissions at the granularity of packages, rather than entire applications or modules within those packages. Application-level permissions can already be assured with traditional sandboxing techniques (e.g. containers), though this would give malicious packages all resources the application may rightfully have. Package-level permissions, as opposed to module-level permissions, is suitable because \emph{packages} are the building blocks that developers install and refer to in their applications, whereas \emph{modules} within a package are rarely referred directly (information hiding). 

We settled on the set of 4 simple permissions after multiple iterations reflecting that they correspond to the most important resources and to allow for flexible and advanced metaprogramming behavior in trusted packages if needed. The approach could be extended with a more fine-grained permission model where access to individual access paths or permissions for specific parameters could be restricted (e.g., file access in a specific directory only), but such approach would raise complexity and runtime overhead. Given that either design would be limited to a subset of modules, we opted for the simpler, more runtime-efficient, and easier to adopt design.

\myparagraph{Integration with user-facing tools}
The integration of the permission system with current tools and developers workflow are an important part for its adoption in practice. Even though we do not evaluate the integration in this paper, we outline the changes in user-facing tools and in developers' workflow that would be required to integrate the permission system in the Node.js/npm ecosystem.

Package owners are expected to manually declare required permissions in the package's manifest file before publishing a package. Package users can see required permissions in the npm repository before installation and will be notified about permission changes and potential permission mismatches at installation/update time. For example, one might be notified that their own package does not have permissions to import another package, so one has to update one's own permissions. Tools like \emph{npm} or \emph{yarn} package managers are expected to \emph{not} automatically update packages when additional permissions are requested in the update, but instead require extra confirmation, encouraging a closer look at the update. 

\myparagraph{Alternative: Taint Tracking}
The most obvious alternative solution would be a policy that restricts
the flow of sensitive information to security-sensitive sinks using
information-flow analysis~\cite{hedin:2016:IFJS,austin:2012:MFDIF,bauer:2015},
such that packages require permissions to initiate certain flows.
Such policy is much more flexible, because rather than completely restricting
access to packages with requiring additional permissions, 
we can allow packages to import other packages as long as they do not
use them (directly or indirectly) to leak sensitive information; it would allow more flexible differentiation
of which information may flow into which sinks independent of which packages
are involved; it might also depend less on the fact
that many packages are simple in their implementation.
In addition, such a policy could be enforced uniformly independently of
how modules obtain references to other modules (actively or passively)
and, given a suitable information-flow tracking system, we would not need
to restrict \emph{eval} and metaprogramming.

However, we ruled out information-flow policies 
because corresponding static information-flow analyses are
notoriously difficult to implement precisely in a dynamic language like 
JavaScript~\cite{guarnieri:2009:GATEKEEPER, staicu:2018:SyNode} and dynamic information-flow analysis tend
to have unacceptable performance overhead to consider adoption in
practice (often 40x-100x performance overhead)~\cite{austin:2012:MFDIF, 25millionflowslater:CCS:2013, jia:ndss:2015:IFCHROMIUM, chudnov:2015:IIF}.
Neither the performance overhead nor a requirement for developers
to rewrite or annotate their code would have made  
ecosystem-wide adoption plausible.

\myparagraph{Alternative: Restricting Inter-Module Communication} 
While our policy restricts the \emph{access} to other packages (and their resources), an alternative
design could assign permissions to modules (as in our design) and then restrict the \emph{communication} across package boundaries:
A corresponding policy could simply state that no code originating from
a module with permissions $X$ shall call code from a module with permissions $Y$ unless $Y\subseteq X$, restricting packages from calling
code from other packages that require more permissions.

Checking function calls to packages is more flexible than checking imports of packages, as it provides further opportunities for inspecting arguments and return values, and observing how packages communicate. 
More importantly, such design could ensure that a package can only call
other packages with suitable permissions even when a reference
to a security-relevant resource is received \emph{passively}, e.g., passed in by the user of the package as argument. Such design could also avoid special
handling of restricted properties that our design requires to avoid unchecked
access to certain references.

Such designs have previously been explored to sandbox individual 
code fragments, usually a single third-party code fragment 
included on a web page~\cite{politz:2011:ADsafety, agten:2012:JSand} or a single module in a Node.js application~\cite{deGroef:2014nodeSentry}.
The key is to track \emph{origins} of all functions (e.g., when a function
is passed from module $A$ to $B$ it still needs to be associated with
$A$) and to intercept all calls that cross module boundaries to check
permissions corresponding to origin modules of the calling and called functions.
In JavaScript, a typical solution is to install monitors between package boundaries using
proxy objects, following the \emph{Membrane Pattern} \cite{cutsem:2013membranePattern, deGroef:2014nodeSentry, tran:2015:JaTE},
but also replacing calls by inter-process communication has been explored~\cite{breakapp:ndss:2018}. 

Despite its elegance, this policy and the corresponding implementations
impose at least three severe challenges that limit its practical use:
% Although this design has the advantage of checking permissions during
% calls and more flexible restrictions on parts of APIs and on specific 
% parameters and return values, we found a number of drawbacks:
\begin{itemize}[leftmargin=*]
	\item The performance overhead of the Membrane Pattern is nontrivial.
Prior work, isolating only a single package, reported about a 20~percent slowdown~\cite{cutsem:2013membranePattern, deGroef:2014nodeSentry, tran:2015:JaTE}; we observed similar and slightly higher slowdowns
in our own experiments installing membranes among all packages.
This overhead may discourage adoption in many settings for
server-side applications.

	\item Node.js packages and many native modules heavily rely on callback functions, for example, to asynchronously return the result of an I/O operation.
Since callback calls also cross package boundaries, it would be subject to the
same permission checks, essentially requiring the same permissions for caller and callee, making
this policy way too restrictive in the context of typical Node.js programming patterns.
Instead, we would have to design a policy that specifically allows certain
call patterns, which is nontrivial and would eventually permit most 
\emph{passive} means of accessing security-critical resources also allowed in our design.

	\item Implementing enforcement with the Membrane Pattern is technically challenging.
	First, while it is easy to install a membrane for modules, using the membrane or other mechanisms to handle
	the \emph{global} object and its properties requires invasive
	changes into Node.js and the underlying JavaScript engine.
	Second, installing proxy objects for objects breaks many existing
	implementations 
	since proxies are not entirely transparent
	(e.g., object identity is no longer reliable).
	We implemented a prototype to run experiments with the Membrane Pattern and had to adapt many existing implementations.
	While technical challenges can potentially be addressed in an automated
	way (invasive changes to Node.js runtime, automated code rewriting to account for proxy behavior), such implementation is complex and difficult to maintain with frequent updates to the Node.js engine.
\end{itemize}
While the policy to restrict inter-module communication seems more elegant
and requires fewer exceptions, performance and implementation challenges 
are too severe to justify its marginal benefits. To account for callbacks
and global, actual policies would likely have similar limitations as our own policy.

\myparagraph{Design space summary}
Overall, while we cannot support the flexibility of a policy
based on taint tracking, our approach provides similar protections
to prior approaches checking inter-module communication, but
with a much simpler implementation and with negligible performance overhead.

While our design cannot provide protections for modules that legitimately
need access to security-critical resources or metaprogramming as part of
their implementation (neither can approaches based on restricting inter-module communication), it protects those small and simple packages that are common in Node.js/npm and do not need any permissions almost for free; Source modules may still need to be inspected, to ensure that they do not provide security-critical resources to target modules with fewer permissions.  

\section{Evaluation}
\label{sec:evaluation}

To evaluate our proposed permission system, we consider the goals of our design (see Sec.~\ref{sec:goalsAndAssumptions}) and show that our permission system (1) can significantly reduce the attack surface of applications and npm in general, (2) is useful in containing real attacks, (3) incurs negligible performance overhead, and (4) can reduce review effort for security purposes on package updates. Specifically, we answer the following research questions:
 \begin{itemize}
	\item 
	\textbf{RQ1:}~How many packages in the \emph{npm} repository could we protect with our permission system?
	\item 
	\textbf{RQ2:}~How effective could the permission system be to contain attacks like the \emph{eslint-scope}, \emph{event-stream}, and \emph{electron-native-notify} attacks?
	 \item 
	\textbf{RQ3:}~What performance overhead would the permission system cause in Node.js applications and is the permission system transparent?
	\item 
	\textbf{RQ4:}~How much review effort could be saved on package updates with our permission system in a realistic setting?
\end{itemize}

The evaluation of our proposed permission system maps directly to our design goals (see Sec.~\ref{sec:goalsAndAssumptions}). Research questions RQ1 and RQ2 map to our first design goal described in Sec.~\ref{sec:goalsAndAssumptions}, which is to propose a solution that \emph{reduces the attack surface} of applications by \emph{containing certain types of attacks}. RQ3 maps to our second and third design goals, which aim at demonstrating transparency and low performance overhead of our proposed permission system design, both important aspects for practical adoption. Finally, RQ4 aims at illustrating another benefit of our (partial) solution: its saves review effort on package updates. It also shows how permission changes are rare and suspicious, especially for minor and patch updates, and that the developer community is much more likely to focus their attention on such updates. 

\subsection{Protected Packages (RQ1)}
\label{sec:evaluation-rq1}
To show how our permission system can help protect packages and applications, significantly reducing the attack surface of applications and npm in general, we approximate the required permission for all packages in the \emph{npm} repository and report how many packages can no longer gain access to security-relevant resources, making malicious updates or other exploits targeting those packages ineffective. To answer RQ1, we report the relative amount of all packages in the \emph{npm} repository that would need each permission (either directly or indirectly through its dependencies).

\myparagraph{Data Collection and Study Design}
To answer RQ1, we use a snapshot of the entire \emph{npm} repository with the latest versions of all packages. We gathered a total of \NRTotalPackages~packages (\DatePackagesSampled{}), from which \NRValidPackages~contain valid JavaScript code. For each package, we analyze both the manifest file and all JavaScript source files (modules) to identify declared dependencies, imports to other modules, and expressions with property accesses. We infer permissions by searching for \emph{require} calls to native modules. We then assign permissions to each package, recursively considering the permissions of its direct and indirect dependencies, as defined by our compositional permission model (see Sec.~\ref{sec:perm-system-design}).

Since not all packages in the \emph{npm} repository are equally used and updated~\cite{decan2019empirical}, we also analyze packages samples of size 100 according to four criteria: most downloaded, most depended on within the \emph{npm} repository, most stars on GitHub (a common popularity measure), and most updated in the year prior to our sample date. We gathered \emph{downloads}, \emph{dependencies}, and \emph{updates} statistics from the \emph{npm} repository and \emph{stars} from GitHub. Note that these criteria look at the dataset from different facets, for example, \emph{downloads} may be biased more toward utility packages used by many other packages, \emph{dependencies} indicate popularity among library developers, \emph{updates} highlight packages that are creating a particularly high review load if one was to review all updates, and \emph{stars} indicate popularity or attention by users more broadly.

Note that our permission inference is an approximation, whereas in practice
we would expect developers to manually declare required permissions. 
Our inference may miss some required permission (that our runtime
enforcement would catch), for example, when imports use dynamically computed names to import native modules
(dynamic imports are found in 8\% of all packages, though they rarely seem to import native modules) and may sometimes infer the \textbf{all} permission for non-problematic access to restricted paths. 
We validate the accuracy of our permission inference by manually identifying permissions of 30 randomly chosen packages and found that those all matched the automatically inferred ones, except 7 cases where we unnecessarily inferred the \textbf{all} permission.
Furthermore, we did not encounter issues from missing permissions in our experiments (Table~\ref{tab:performance-results}), further indicating
accurate inference.

\begin{table}[t]
	\centering
	\small
	\begin{tabular}{lrrrrr} \toprule
		{Permissions} & {all} & {downl.} & dep. & stars & upd.\\ 	\hline
\textbf{no} perm. & 31.9\% & 27\% & 15\% & 14\% & 33\%\\
only \textbf{network} perm. & 1.2\% & 0\% & 1\% & 0\% & 3\%\\
only \textbf{filesys.} perm. & 4.8\% & 8\% & 11\% & 1\% & 6\%\\
only \textbf{process} perm. & 0.7\% & 0\% & 1\% & 0\% & 0\%\\
multiple perm. & 2.7\% & 5\% & 2\% & 0\% & 2\%\\
\textbf{all} perm. & 58.7\% & 60\% & 70\% & 85\% & 56\%\\
		\bottomrule
	\end{tabular}
	\caption{Distribution of permissions of all packages and 
	the 100 most downloaded/depended/starred/updated packages.}
	\vspace{-1em}
	\label{tab:evaluation-permissions} 
\end{table}

\myparagraph{Results}
Based on our permission inference, we found that \NoPermissionsPackages{} packages in the \emph{npm} repository (\AttackSurfReduction{}) do not need any permission 
and another 9.4\% need only a subset of all permissions. Among the most downloaded, depended, starred, and updated packages, a higher percentage of packages needs some or all permissions, but still a significant number of those packages need no (14--33\%) or only some permissions (1--11\%) as shown in Table~\ref{tab:evaluation-permissions}. The differences in different populations are to be expected, as popular (starred) packages tend to be larger end-user packages, whereas downloads favor smaller, often indirectly used utilities.

Note, many packages require permissions primarily due to dependencies: We infer that 21.9\% of all packages directly need the \textbf{all} permission, while another 36.8\% inherit the \textbf{all} permission from depended on packages. Among packages that directly need the \textbf{all} permission, around 5\% directly use \emph{eval} while the remaining of packages change the prototype of native objects such as \emph{Object}, \emph{String}, \emph{Array}, and others. Similarly, we infer that only 4.7, 15.6, and 2.1\% of all packages need the \textbf{network}, \textbf{filesystem}, and \textbf{process} permissions. Multiple permissions refer to combinations of permissions (e.g., \textbf{network}+\textbf{process}) and packages that need them are rarely found in the \emph{npm} repository.

Overall, while there are a large number of packages that genuinely need access to security-critical APIs and a significant number of packages that use language features that cannot be contained by our mechanism (hence the \textbf{all} permission), these results also confirm that many packages published on the \emph{npm} repository are indeed fairly simple and can be protected with our lightweight permission and enforcement system. Since permissions are enforced for all packages (including their clients), any attempt of a package to import a security-relevant resource that does not correspond to its permissions would fail, even if a package uses a transitive dependency relationship with another package. 

% \noindent
\emph{In summary, our permission system can protect the 14--33\% packages that need no permission and partially protect another 1--11\% percent that need only a subset of all permissions. These packages can no longer gain access to security-relevant resources, making malicious updates that attempt to elevate packages' privileges ineffective. Thus, reducing the attack surface for the npm repository and for typical Node.js applications that use these packages.}

\label{sec:evaluation-rq2}
\subsection{Containing Past Attacks (RQ2)}
Our permission system is a general defense mechanism, but is not designed to detect new malicious attacks in the wild. It is designed to contain attacks. We illustrate the usefulness of the proposed permission system by containing some past attacks. To that end. we replay the \emph{eslint-scope}, \emph{event-stream}, and \emph{electron-notify-native} attacks using our modified Node.js engine and show that the attacks would be ineffective.

\myparagraph{Data Collection and Study Design}
To evaluate the containment, we considered two versions of each attacked package: (i)~the version that preceded the attack and (ii)~the attacked version. First, we inferred the permissions of the version that preceded the attack and assigned the same permissions to the attacked version. Then, we installed the attacked version with the \emph{npm install} command, which installs package dependencies and executes the installation scripts of the package. To replay the attacks on the \emph{eslint-scope}, \emph{event-stream}, and \emph{electron-notify-native} packages, we used versions 3.7.2; 0.1.1 of the \emph{flatmap-stream} package; and 1.1.6, respectively.
\looseness=-1

\myparagraph{Results}
None of the packages with malicious updates required any permissions in their latest release
before the attack.
When replaying the attacks, we observed in each case that the permission system contains all three attacks by preventing unauthorized imports of security-relevant resources and by denying the use of metaprogramming constructs at runtime.
For example, when the malicious version 3.7.2 of the \emph{eslint-scope} package is installed, a \emph{post-install} script is executed, but the permission system prevents the imports of the \emph{http} and \emph{fs} modules. This occurs because the malicious version 3.7.2 did not have the corresponding permissions. An attacker would have had to explicitly request additional permissions, which makes the detection of the attack much more likely.

% \noindent
\emph{In summary, our permission system can successfully contain past malicious updates.}

\label{sec:evaluation-rq3}
\subsection{Performance and Transparency (RQ3)}
To evaluate practicality, which is important for adoption (see Sec.~\ref{sec:goalsAndAssumptions}), we measure the performance overhead caused by the enforcement mechanism for our policy (see Sec.~\ref{sec:perm-system-design-enforc}) and whether it transparently supports 
the execution of packages without modifications (assuming sufficient permissions). 

To be realistic, we measure the runtime overhead of \emph{applications}, which represent actual usage scenarios relevant for end users, rather than conducting microbenchmarks on individual packages. To answer RQ3, we evaluate and report the performance overhead caused for \NrApplicationsSampled{} Node.js applications when executing them with and without the permission system. 

\myparagraph{Data Collection and Study Design}
As subject systems, we collected 20~command-line applications and corresponding realistic workloads; which are common modern use cases 
for Node.js and released as open source (e.g., whereas for web services most public examples are demos or tutorials, most production servers are closed source).
We selected the 20~most popular \emph{npm} public applications that depend on the commonly used \emph{commander} package, which provides command-line interfaces for Node.js applications. For each application, listed in Table~\ref{tab:performance-results}, we read the documentation and identified common inputs. Then, we selected a realistic large workload (usually a large JavaScript/JSON file) and measured performance when executing each application with an input.

To mitigate {systematic errors} in our performance measurements, we followed standard guidelines~\cite{georges:2007:rigorousJavaPerformanceEvaluation}, running each application 10 times under each experimental condition, interleaving the runs across applications, and reporting the median execution time. To avoid the noise problems of microbenchmarks, we selected large workloads taking at least five seconds. All measurements were performed on a MacBook Pro equipped with a 2.8GHz Intel Core i7 processor and 16GB 1600MHz DDR3 of memory. The benchmarks are also shared with the project's repository.

\begin{table}[t]
	\small
	\centering
	\setlength\tabcolsep{5pt}
	\begin{tabular}{lrrrrrr}
		\toprule
		{Application} & {Baseline} & {Overhead}& {Import}&  {Property} \\
		&{ (in ms)}& { (in \%)}& {Checks}& {Checks} \\
\midrule
d3-dsv &   5268.0 &   0.3 &             25 &                   408 \\
docco & 5148.5 &   0.0 &             71 &                 65774 \\
dot-object &   5339.5 &   0.2 &             40 &                     2 \\
dox &   5408.5 &   0.0 &           3478 &                    21 \\
findup &   5227.5 &   0.1 &             12 &                     0 \\
html-minifier &  5700.5 &   0.0 &             44 &                     3 \\
js-cfb&   5300.0 &   0.0 &             10 &                     0 \\
js-xss&   5261.5 &   0.3 &             11 &                     0 \\
js-yaml-front-matter &  5200.0 &   0.0 &              7 &                     7 \\
json-refs &   5894.5 &   0.0 &             88 &                   750 \\
json2csv &  5317.0 &   0.1 &             15 &                     3 \\
juice&   6143.0 &   -0.3 &            301 &                  1104 \\
metalsmith &   5828.5 &   0.3 &            148 &                   127 \\
mocha &    5141.5 &    0.0 &            116 &                     5 \\
mock &  4947.0 &   1.3 &              7 &                     0 \\
node &   6627.5   &0.2 &            934 &                   106 \\
sails &   6792.0 &   0.2 &           1364 &                 36455 \\
svgicons2svgfon &   5427.5 &   0.1 &             57 &                     2 \\
traceur-compiler &   13784.0 &  0.5 &             52 &                     2 \\
uglify-js &  12468.0 &  0.9 &             10 &                     1 \\
\bottomrule
	\end{tabular}
	\caption{Applications and performance results; runtime of the unmodified Node.js (Baseline) and the overhead of the modified Node.js with the permission system (Overhead); reporting also the number of executed import and property access checks.}
	\vspace{-1em}
	\label{tab:performance-results} 
\end{table}

\myparagraph{Results}	
We report the observed performance differences in Table~\ref{tab:performance-results}, which across all applications, 
is negligible. The permission system causes a small performance overhead---barely distinguishable from measurement noise---typically under 1~percent (average $0.2\%$). The difference in overhead between the executions with the unmodified and the modified Node.js is not significant (Welch's $t$(0.0218) = 37.9964, $p$ < 0.9827).
\looseness=-1

Notice how the overhead caused by the permission system is negligible, compared to alternative isolation and compartmentalization solutions~\cite{breakapp:ndss:2018, deGroef:2014nodeSentry, tran:2015:JaTE} with $>20\%$ overhead or information flow solutions with $> 20\times$ overhead; cf. Sec.~\ref{sec:alternativeDesign}). 
The number of dynamic import checks and 
property checks needed in those applications (including all
direct and indirect dependencies) differ substantially
from application to application depending on how many and what kind of
dependencies they use, but they are usually low, explaining the low
runtime overhead. For instance, \emph{docco} and \emph{sails} are both HTML generators. While the amount of \emph{actually executed property checks} is much larger for these two applications (as we show in Table~\ref{tab:performance-results}), the amount of \emph{rewritten property checks} is not. For these two applications, the rewritten property checks get executed several times when generating HTML. By inspecting their code, we observed they change the prototype of several native objects (e.g., \emph{String}, \emph{Object}), which cause the rewrites.

Even in systems with higher numbers of property access checks,
the overhead of these checks is dwarfed by the main computations
and I/O operations of these applications.

Furthermore, our experiments confirm that our solution is entirely transparent for the executed applications, if permissions are correctly set. All applications work as before without
any source code modification of the application or its dependencies.

%The result of our t-test between the ten executions of the unmodified Node.js and the modified Node.js with the permission system (import checks and property access checks) is t(\performanceReqMembDOF{}) = \performanceReqMembTtest{}, p = \performanceReqMembPvalue{}.

\emph{In summary, our permission system is transparent and causes negligible performance overhead ($\ll 1\%$).}

\label{sec:evaluation-rq4}
\subsection{Review Effort Reduction on Package Updates (RQ4)}
Our final research question hypothesizes a scenario in which developers
would review all package updates (which may be recommended but is
rarely done systematically in practice, cf.\ Sec.~\ref{sec:stateoftheart-defense}).
We argue that even though, in practice, developers do not review all updates, 
our permission system would allow developers to save even more effort 
if they skip reviews for those packages that do not need any permissions and 
pay particular attention to updates that request additional permissions. 
Thus, we evaluate how our permission system can reduce the review
load in this scenario, indirectly demonstrating the potential
usefulness of our (partial) solution.

\myparagraph{Data Collection and Study Design}	
We analyze package updates and observe how permissions evolve by
replaying the evolution and updates of dependencies for a sample
of packages and applications over a one-year period.
We observe updates both from the perspective of a package maintainer of highly depended packages
and from the perspective of an application developer by analyzing two datasets: (i) the 100-most-depended-upon packages in the \emph{npm} repository that we previously described for RQ1 and (ii) the 20 Node.js applications from RQ3. For each of the 120 analyzed packages and applications, we computed their entire dependency tree (including resolving the latest version
of a dependency when version constraints are declared as a range)
for each day in 2018 and identified
all package updates that happened in that time period.
This is equivalent to installing or updating these applications
every day, approximating common practice among
Node.js developers~\cite{bogart:2016BreakingAPIs}.

We downloaded all distinct packages and their versions to infer
approximate permissions (as in RQ1). 
For each update, we collect whether and how the corresponding
permissions have changed. In total, we analyzed 4,962~distinct dependency versions for the 100~most-depended-upon packages and 1,310~distinct dependency versions for the 20~applications.

\myparagraph{Results}
During the one-year period, a total of 5,042 package updates occurred for all direct and indirect dependencies of the 100-most-depended-upon packages; an average of 66 updates per year per subject.
Among these, 2,644 updates (52 percent) were to packages that did not need any permissions before or after and could be installed without any review -- this represents a substantial reduction in review effort in our scenario.
The percentage of no-permission updates naturally differed among the datasets, but represented a substantial number of updates in most cases.
Applications, on average, had a lower proportion of no-permission updates (6~percent), explained by the larger number of non-trivial packages they depended on. Updates from packages with all permissions make up 33 and 86~percent of all updates in the top libraries and applications respectively.

\begin{figure}[t]
	\centering
	\includegraphics[width=\linewidth]{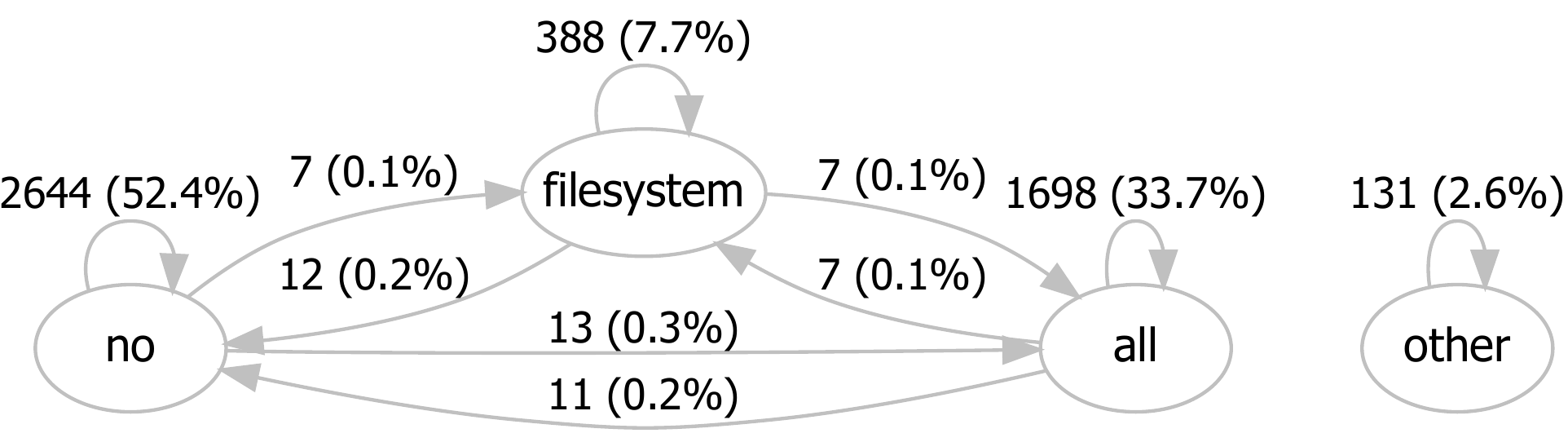}
	\caption{Permission changes for the 100-most-depended-upon packages and all its direct and indirect dependencies. \emph{Nodes} represent permissions, \emph{edges} represent permission changes in updates, and labels indicate the absolute and relative number of changes over a year.}
	\label{fig:perm-changes-graph}
	\vspace{-1em}
\end{figure}

Changes among the permissions in updates in the top libraries were generally rare,
as shown in Figure~\ref{fig:perm-changes-graph}. There are only 27
updates (0.5~percent) that needed more permissions than the previous
one; most maintainers would not face such an update more than once a year,
making it realistic that they would analyze such packages more closely. Nodes that represent other permissions (e.g., the \emph{network} node) and and edges that represent other permission changes (e.g., the edge from \emph{filesystem} to the \emph{network}) are grouped in the \emph{other} node and account for only 2~percent of the updates (see Figure \ref{fig:perm-changes-graph}).

\emph{In summary, our permission system can reduce review effort for updates substantially (6--52\%) in our scenario.}

\subsection{Threats to Validity}
First, different subpopulations of packages in the \emph{npm} repository may have different
characteristics and may need more or fewer permissions on average;
our results must be interpreted as averages over all of packages and
averages over popular packages in the \emph{npm} repository.
Second, it is challenging to assemble a representative set of benchmarks of \emph{Node.js} applications: while libraries are published on the \emph{npm} repository, applications are often proprietary and not public on the \emph{npm} repository or \emph{GitHub}. We use publicly released command-line utils to represent applications, but generalizations must be made with care.
Third, as discussed, our permission inference is only an approximation of 
the permissions a developer would manually declare; despite our careful validation
we may miss permissions or developers may choose to over-permission their
packages affecting the results.
Finally, the evaluation of review effort makes strong assumptions on developer behavior and uniform review effort per package that may not be realistic in practice, especially with often
observed complacency or a false sense of security from a security mechanism like
ours; readers may extrapolate other behavior from the reported numbers.

\section{Related Work}
\label{sec:related-work}
Security challenges with the JavaScript programming language and the limits of JavaScript analysis techniques have been widely explored in the web browser context, with less focus on the Node.js/npm ecosystem.

\myparagraph{Node.js/npm Ecosystem} \citet{ojamaaDuuna:2012:NodeJS} discussed security challenges of the Node.js platform, pointing out typical pitfalls of JavaScript programming, such as the single-threaded event-loop-based architecture, \emph{eval}, lack of isolation, but also highlighting the possibility of attacks through malicious package updates. Some of the pitfalls (e.g., \emph{eval}) could result in insecure applications and are partially addressed by our permission system. \citet{wittern:2016} found that the number of direct dependencies of Node.js packages grows over time and that over 40\% of all packages allow automatic updates of minor revisions. These results reinforce the common but problematic practice of accepting automatic package updates (see Sec. \ref{sec:characteristics}). \citet{decan:2017} reported that the \emph{npm} repository has an abundant number of packages whose failure can impact the ecosystem, some affecting more than 30 percent of the packages available on the \emph{npm} repository. \citet{abdalkareem:2017} studied the pervasive use of small, single purpose packages in the \emph{npm} repository, which motivated our permission system focused on the many simple packages. 
\looseness=-1

\citet{hejderup:2015:masterThesis} analyzed dependencies among packages published on the \emph{npm} repository and found that known vulnerabilities in packages often affect many other dependent packages in the ecosystem, with many packages  depending on vulnerable versions for a significant time after a patch has been released. Similarly, \citet{decan:2018:vulnerabilitiesNPM} analyzed packages in the \emph{npm} repository over six years and found that package vulnerabilities affect many dependent packages, but it still takes a long time for them to be discovered and fixed. \citet{zimmermann:2019:smallWorldHighRisks} further reported many unmaintained packages in the \emph{npm} repository, which no longer receive patches, indirectly threatening the security of the Node.js/npm ecosystem. \citet{staicu:2018:SyNode} further reported how injection vulnerabilities are prevalent in the Node.js/npm ecosystem. These results highlight an orthogonal but relevant issue: not updating vulnerable package dependencies can cause as much damage as accepting malicious package updates. In practice, many developers rely on tools to track known vulnerabilities (Sec.~\ref{sec:stateoftheart-defense}). 

\myparagraph{JavaScript Analysis} 
It is well understood that analyzing and sandboxing JavaScript code is difficult. Prior work, usually focused on untrusted scripts embedded in web pages, restricts how third-party code can interact with application code, a browser, or the web page. Due to the dynamic nature of JavaScript programs, many existing analysis approaches combine static analysis with runtime mechanisms, like our approach. Strategies typically either define and enforce a safe subset of JavaScript \cite{crockford:2008:adsafe, maffeis:2009:LBIUJS, finifter:2010:capabilityLeaksJS, politz:2011:ADsafety, kashyap:2014:JSAI, bhargavan:2014:defensiveJS}, isolate scripts/packages in different execution environments (e.g., \emph{iframes}) \cite{stefan:2014, ingram:2012:treehouse, mickens:2014:pivot, louw:2010:AdJail}, modify the browser to enforce restrictions at runtime \cite{meyerovich:2010:ConScript}, or attempt to sandbox a script/package without JavaScript engine modifications by rewriting it \cite{agten:2012:JSand, deGroef:2014nodeSentry, tran:2015:JaTE, sun:2018:efficientDynAnalysisNodeJS, breakapp:ndss:2018}.

We combine several of these ideas (language subset, rewriting, runtime modification) in a lightweight design.
More precise dynamic information flow techniques have been widely investigated in academic literature, but usually require very invasive changes to execution engines and are very slow \cite{austin:2012:MFDIF, chudnov:2015:IIF, hedin:2016:IFJS, hedin:2014:JSFLow}; static information flow analysis is only feasible with requiring developers to write fine-grained annotations \cite{arden:2012:mobileCodeIFC}. 

\myparagraph{Permission Systems and Sandboxing} Permission systems are common to provide controlled access to security-relevant resources, allowing users to monitor, review, and revoke applications' permissions, if applications' behaviors (or intentions) are misaligned with application users' expectations. Our permission system is inspired by the one available in Android OS, which has been broadly studied \cite{felt:2011:effectiveAppPermissions, felt:2011:androidPermDemystified, wei:2012:permEvolutionAndroid, bartel:2012:reducingAttackSurfaceAndroid, holavanalli:2013:flowPermisionsAndroid}, which isolates apps from each other rather than packages within a single application. While the systems are not directly comparable and have different user populations, we can learn from their experience, including well studied problems such as developers asking for too many permissions and users ignoring permissions~\cite{bartel:2012:reducingAttackSurfaceAndroid, wei:2012:permEvolutionAndroid, felt:2011:androidPermDemystified, guha:2011:verifBrowserExt}. 

Sandboxes are important building materials to harden the security of diverse software systems and are typically used simultaneously as (i) encapsulation and as (ii) policy enforcement mechanisms \cite{maass:2016}. In our case, we effectively implement a sandbox in the Node.js engine, to enforce our permission-dependent policy to limit the  behaviors of packages at runtime. More commonly, sandboxing isolates programs in the operating system or in containers \cite{zeng:2014,wan2019practical,shinagawa2009bitvisor}, or, for JavaScript, isolates untrusted code in the browser \cite{crockford:2008:adsafe, politz:2011:ADsafety, agten:2012:JSand}.

\section{Conclusion}
\label{sec:conclusion}
In this paper, we discuss the emerging security challenges of malicious package updates which recently surfaced in multiple software ecosystems~\cite{eslint:2018:postmortem, getcookies:2018:postmortem, event-stream:2018:postmortem, electron-native-notify:2019:postmortem, maliciousPackagesRuby, maliciousPackagesPython}.
We design a permission system with a policy and corresponding enforcement
mechanism that sandboxes individual packages, rather than entire applications, ensuring that malicious updates cannot use security-critical
resources for which they do not have permissions.
We design our system to be simple, easy to adopt, with marginal
runtime overhead and show that 
\AttackSurfReduction\ of all packages can be protected at almost no cost.

\section*{Acknowledgment}
This work has been supported by NSF award 1717022. G. Ferreira is supported by a doctoral research grant from CAPES, Brazil.

\bibliographystyle{ACM-Reference-Format}
\small
\bibliography{icse2021-cameraready.bib}

\end{document}